\documentclass{aa}

\usepackage{graphicx, psfig, times, amsfonts}

\newcommand{\less}{\raisebox{-1.1mm}{$\stackrel{<}{\sim}$}}

\begin{document}

\title{The metallicity dependence of the Cepheid $PL$-relation} 

\author{
M.A.T.~Groenewegen
\inst{1} 
\and
M.~Romaniello
\inst{2} 
\and
F.~Primas
\inst{2} 
\and
M.~Mottini
\inst{2} 
}

\institute{
Instituut voor Sterrenkunde, PACS-ICC, Celestijnenlaan 200B, 
B--3001 Leuven, Belgium 
\and
ESO, Karl Schwarzschild stra{\ss}e 2, D-85748 Garching, Germany 
}

\date{received: 2003,  accepted: 20 February 2004}

\offprints{Martin Groenewegen (groen@ster.kuleuven.ac.be)}


\abstract{
A sample of 37 Galactic, 10 LMC and 6 SMC cepheids is compiled for
which individual metallicity estimates exist and $BVIK$ photometry in
almost all cases. The Galactic cepheids all have an individual
distance estimate available. For the MC objects different sources of
photometry are combined to obtain improved periods and mean
magnitudes. A multi-parameter Period-Luminosity relation is fitted to
the data which also solves for the distance to the LMC and SMC.  When
all three galaxies are considered, without metallicity effect, a
significant quadratic term in $\log P$ is found, as previously
observed and also predicted in some theoretical calculations. For the
present sample it is empirically determined that for $\log P < 1.65$
linear $PL$-relations may be adopted, but this restricts the sample to
only 4 LMC and 1 SMC cepheid.  Considering the Galactic sample a
metallicity effect is found in the zero point in the $VIWK$
$PL$-relation ($-0.6 \pm 0.4$ or $-0.8 \pm 0.3$ mag/dex depending on
the in- or exclusion of one object), in the sense that metal-rich
cepheids are brighter. The small significance is mostly due to the
fact that the Galactic sample spans a narrow metallicity range. The
error is to a significant part due to the error in the metallicity
determinations and not to the error in the fit. Including the 5 MC
cepheids broadens the observed metallicity range and a metallity
effect of about $-0.27 \pm 0.08$ mag/dex in the zero point is found in
$VIWK$, in agreement with some previous empirical estimates, but now
derived using direct metallicity determinations for the cepheids
themselves.
}


\maketitle

\section{Introduction}

The importance of the cepheid Period-Luminosity relation
($PL$-relation) has long been recognised and is the basis of the
determination of the Hubble constant by Mould et al. (2000) and
Freedman et al. (2001). The most important uncertainties in this
derivation are the zero point of the $PL$-relation based on an adopted
distance modulus (DM) to the LMC, and the adopted metallicity
correction.  Mould et al. (2000) and Freedman et al. (2001) adopt
corrections in the Wesenheit-index ($W$ = $V - 2.55 (V-I)$) of $-0.24
\pm 0.16$ and $-0.2 \pm 0.2$ mag/dex, respectively. This metallicity effect
is important as the galaxies surveyed by the HST Key Project span a
factor of 30 in oxygen abundance (Ferrarese et al. 2000).

Theoretical pulsation models lead to different results: linear models
(e.g. Sandage et al. 1999, Alibert et al. 1999, Baraffe \& Alibert
2001) predict a moderate dependence, while non-linear convective
models (Bono et al. 1999, Caputo et al. 2000) predict a larger
dependence, and in the sense that metal-rich cepheids are fainter (see
Table~7 in Groenewegen \& Oudmaijer 2000, for values at typical
periods).  Recently, Fiorentino et al. (2002) suggested that there is
also a dependence on the Helium abundance.

On the observational side the results seem to indicate consistently
that metal-rich cepheids are brighter, and various estimates have been
given in the literature, $-0.88 \pm 0.16$ mag/dex ($BRI$ bands, Gould
1994;), $-0.44^{+0.1}_{-0.2}$ mag/dex ($VR$ bands, Sasselov et
al. 1997), $-0.24 \pm 0.16$ mag/dex ($VI$ bands, Kochanek 1997),
$-0.14 \pm 0.14$ mag/dex ($VI$ bands, Kennicutt et al. 1998), $-0.25
\pm 0.05$ mag/dex ($VI$ bands, Kennicutt et al. 2003), and $-0.21 \pm
0.19$ in $V$, $-0.29 \pm 0.19$ in $W$,$-0.23 \pm 0.19$ in $I$,$-0.21
\pm 0.19$ mag/dex in $K$ (Storm et al. 2004).  The potential drawback
or caveat is that no individual abundance determinations of individual
cepheids are being used in these studies but rather abundances of
nearby H{\sc ii} regions, or even a mean abundance of the entire
galaxy.

The present paper aims at investigating the metallicity dependence
from the observational side, but using cepheid individual metallicity
determinations. This has become possible because of advances in
abundance determinations for the Galaxy (Fry \& Carney 1997,
Andrievsky et al. 2002a,b,c, Luck et al. 2003) and LMC (Luck et
al. 1998), as well as recent advances in individual distance estimates
for Galactic cepheids based on surface-brightness relations (Fouqu\'e
et al. 2003), a Bayesian statistical analysis to solve the
surface-brightness equations (Barnes et al. 2003), direct measurement
of distances based on combining radial velocity data with
interferometric observations (Kervella et al. 2003), and distance
determinations of cepheids in open clusters (Tammann et al. 2003).

The paper is organised as follows. In Sect.~2 the datasets on
individual distance and metallicity determinations for Galactic
cepheids are presented and compared. The Magellanic Cloud sample is also
described. In Sect.~3 the model and the results are presented. The
conclusions and future prospects are outlined in Sect.~4.

\section{The data}

\subsection{Galactic cepheids}

Distances for Galactic cepheids have been taken from Benedikt et
al. (2002, B02), Nordgren et al. (2002, N02), Lane et al. (2002,
L02), Fouqu\'e et al. (2003, F03), Barnes et al. (2003, B03),
Kervella et al. (2003, K03), and Tammann et al. (2003, T03). Some of
them have independent determinations, and these are compared in
Table~\ref{TAB-GAL-D}. T03 (largely based on Feast 1999) does not
quote error bars and here a uniform error bar of 0.15 in DM is
assigned as recommended by Feast (1999), that takes into account the
error in the Pleiades distance, as well as the uncertainty due to
the reddening in the main-sequence fitting method which is the method
used for these distances.  The agreement between the different
determinations is (surprisingly) good in many cases. The finally
adopted distances (last column in Table~\ref{TAB-GAL-D}) are weighted
means, but with the error bar for the F03 determinations multiplied by
a factor of 5.  The error bars quoted by F03 appear very small when
compared to those in B03 which are also based on the surface-brightness 
technique, and sometimes appear even smaller than the rms error in the
fundamental surface-brightness relations (e.g. Nordgren et al. 2002). 
The factor of 5 roughly brings the error bars in F03 to the level of
B03. When the individual determinations are very different the error
bar is slightly increased (WZ Sgr, $l$ Car).

\begin{table*}
\caption{Comparison of distances (in parsec) for Galactic cepheids with independent determinations}
\begin{tabular}{rrrrrrrrr} \hline

Name       &       F03     &      K03     &        B03     &      T03       &       B02      &       N02   &      L02     & Adopted (pc) \\
\hline
$\delta$ Cep& 261 $\pm$  5 &              &                &                &  271 $\pm$  11 & 272 $\pm$ 6 &              &  271 $\pm$   5 \\
$\zeta$ Gem&               & 360 $\pm$ 25 &                &                &                &             & 362 $\pm$ 38 &  360 $\pm$  20 \\
$\eta$ Aql &  250 $\pm$  6 & 261 $\pm$ 14 &                &                &                &             & 320 $\pm$ 32 &  267 $\pm$  12 \\
$l$ Car    &  628 $\pm$  3 & 524 $\pm$ 49 &                &                &                &             &              &  620 $\pm$  30 \\
X Cyg      & 1214 $\pm$  9 &              & 1101 $\pm$ 28  &                &                &             &              & 1130 $\pm$  25 \\
T Mon      & 1430 $\pm$ 35 &              & 1306 $\pm$ 41  & 1690 $\pm$ 120 &                &             &              & 1350 $\pm$  40 \\
BF Oph     &  715 $\pm$ 11 &              &  713 $\pm$ 63  &                &                &             &              &  714 $\pm$  40 \\
RS Pup     & 2111 $\pm$ 73 &              & 1706 $\pm$ 228 & 1800 $\pm$ 130 &                &             &              & 1810 $\pm$ 110 \\
U Sgr      &  595 $\pm$  6 &              &  672 $\pm$ 49  &  643 $\pm$  55 &                &             &              &  620 $\pm$  25 \\
WZ Sgr     & 1808 $\pm$ 40 &              & 2513 $\pm$ 196 & 1790 $\pm$ 125 &                &             &              & 1960 $\pm$  90 \\
BB Sgr     &  801 $\pm$ 10 &              &  914 $\pm$ 76  &  664 $\pm$  47 &                &             &              &  760 $\pm$  30 \\
RY Sco     & 1268 $\pm$ 20 &              &  960 $\pm$ 65  &                &                &             &              & 1050 $\pm$  55 \\
SZ Tau     &               &              &  560 $\pm$ 85  &  555 $\pm$  40 &                &             &              &  556 $\pm$  35 \\ 
SV Vul     &               &              & 1846 $\pm$ 69  & 2320 $\pm$ 170 &                &             &              & 1910 $\pm$  70 \\
CV Mon     & 1576 $\pm$ 25 &              &                & 1755 $\pm$ 300 &                &             &              & 1600 $\pm$ 115 \\
V Cen      &  684 $\pm$ 20 &              &                &  682 $\pm$  49 &                &             &              &  684 $\pm$  45 \\
S Nor      &  959 $\pm$ 14 &              &                &  933 $\pm$  67 &                &             &              &  945 $\pm$  50 \\
V340 Nor   & 1690 $\pm$ 155 &             &                & 1713 $\pm$ 125 &                &             &              & 1700 $\pm$ 100 \\
VY Car     & 2000 $\pm$ 20 &              &                & 2118 $\pm$ 150 &                &             &              & 2040 $\pm$  80 \\
RZ Vel     & 1600 $\pm$ 20 &              &                & 1730 $\pm$ 120 &                &             &              & 1650 $\pm$  80 \\
SW Vel     & 2510 $\pm$ 30 &              &                & 2610 $\pm$ 180 &                &             &              & 2550 $\pm$ 115 \\
U Car      & 1560 $\pm$ 25 &              &                & 1960 $\pm$ 140 &                &             &              & 1740 $\pm$  90 \\
\hline
\end{tabular}
\label{TAB-GAL-D}
\end{table*}

The absolute magnitudes in $BVIK$ have been taken from the following
sources in order of preference. F03, where appropriate, changed to
allow for the finally adopted distance in Table~\ref{TAB-GAL-D}, or
for the increased error bar in the distance. T03 for the $BVI$
absolute magnitudes (where appropriate, changed to allow for the
finally adopted distance), with observed $K$ magnitudes from
Groenewegen (1999), dereddened using the $E(B-V)$ listed by T03, and
using $A_{\rm K}$ = 0.3 $E(B-V)$. For the stars only listed in B03,
the observed $BVIJK$ magnitudes come from Groenewegen (1999) and from the
electronic database of Fernie et al. (1995). Recent $K$-band
photometry for CF Cas and DL Cas was taken from Hoyle et
al. (2003). Dereddening was done using the values in the electronic
database using $A_{\rm V}$ = 3.3 $E(B-V)$, and $A_{\rm B}$ = 1.33
$A_{\rm V}$, $A_{\rm I}$ = 0.60 $A_{\rm V}$, $A_{\rm K}$ = 0.091
$A_{\rm V}$.

Abundance determinations have been considered from the recent papers
by Fry \& Carney (1997), Andrievsky et al. (2002a, b, c) and Luck et
al. (2003)\footnote{For Andrievsky et al. (2002a, c) and Luck et
al. (2003) the errors in the abundances are listed in the
electronically available tables.}. A comparison of stars in common is
given in Table~\ref{TAB-GAL-Z}, which also lists the adopted value in
this study. The disagreement between the studies is at times
worrisome, and the impact will be discussed in Sect.~4. It also
appears from the work of Andrievsky et al. and Luck et al., as well as
from the intercomparison made in Table~\ref{TAB-GAL-Z} that the errors
listed in Fry \& Carney are unrealistically small, and for stars
that only appear in that paper a minimum error in [Fe/H] of 0.06 dex
has been adopted\footnote{This error is also more realistic
considering the error in the derived metallicity due to uncertainties
in the effective temperature and gravity, as Fry \& Carney themselves
discuss in their Table~7.}.

\begin{table*}
\caption{Comparison of [Fe/H] values for Galactic cepheids with independent determinations}
\begin{tabular}{rrrrr} \hline

Name         & Fry \& Carney (1997) & Andrievsky et al. (2002a) & Luck et al. (2003) & Adopted \\
\hline
V340 Nor     & $-0.18 \pm 0.03$ & $-0.00 \pm 0.10$ &                  & $-0.09 \pm 0.08$ \\
CF Cas       & $-0.20 \pm 0.02$ & $-0.01 \pm 0.09$ &                  & $-0.10 \pm 0.07$ \\
DL Cas       & $+0.05 \pm 0.01$ & $-0.01 \pm 0.14$ &                  & $+0.02 \pm 0.10$ \\
$\delta$ Cep & $-0.01 \pm 0.06$ & $+0.06 \pm 0.07$ &                  & $+0.03 \pm 0.06$ \\
CV Mon       & $-0.05 \pm 0.06$ & $-0.03 \pm 0.12$ &                  & $-0.04 \pm 0.06$ \\
V Cen        & $-0.14 \pm 0.02$ & $+0.04 \pm 0.09$ &                  & $-0.05 \pm 0.06$ \\
U Sgr        & $+0.01 \pm 0.03$ & $+0.04 \pm 0.08$ &                  & $+0.03 \pm 0.06$ \\
$\eta$ Aql   & $+0.07 \pm 0.04$ & $+0.05 \pm 0.06$ &                  & $+0.06 \pm 0.06$ \\
S Nor        & $-0.03 \pm 0.02$ & $+0.05 \pm 0.08$ &                  & $+0.01 \pm 0.07$ \\
X Cyg        & $+0.12 \pm 0.03$ & $+0.12 \pm 0.05$ &                  & $+0.12 \pm 0.05$ \\
WZ Sgr       & $-0.15 \pm 0.03$ & $+0.17 \pm 0.08$ &                  & $+0.00 \pm 0.15$ \\
SW Vel       & $-0.08 \pm 0.03$ & $+0.01 \pm 0.08$ & $-0.07 \pm 0.09$ & $-0.05 \pm 0.06$ \\
T Mon        & $+0.09 \pm 0.02$ & $+0.13 \pm 0.06$ & $+0.13 \pm 0.12$ & $+0.11 \pm 0.06$ \\
$\zeta$ Gem  & $+0.00 \pm 0.04$ & $+0.04 \pm 0.07$ &                  & $+0.02 \pm 0.05$ \\
RX Aur       & $-0.13 \pm 0.04$ & $-0.07 \pm 0.08$ &                  & $-0.10 \pm 0.07$ \\
SZ Tau       & $-0.01 \pm 0.03$ & $+0.08 \pm 0.07$ &                  & $+0.04 \pm 0.06$ \\
T Vul        & $-0.10 \pm 0.04$ & $+0.01 \pm 0.05$ &                  & $-0.05 \pm 0.06$ \\
SV Vul       & $+0.06 \pm 0.04$ & $+0.03 \pm 0.09$ &                  & $+0.05 \pm 0.06$ \\
\hline
\end{tabular}
\label{TAB-GAL-Z}
\end{table*}

The finally adopted sample of stars with recent distance and
metallicity determinations is listed in Table~\ref{TAB-GAL-P}, and
consists of 37 objects. Listed are the name, observed period, adopted
distance modulus with error, corresponding absolute magnitudes in
$BVIK$, adopted reddening and metallicity with error, and finally a
code for fundamental mode (FU = 0) anf first overtone (FO = 1) pulsators.

\begin{table*}
\caption{Basic data of Galactic cepheid sample with individual distances and metallicity determinations}
\begin{tabular}{rrrrrrrrrr} \hline

Name       & $\log P$ &     DM       & $M_{\rm B}$ & $M_{\rm V}$ & $M_{\rm I}$ & $M_{\rm K}$ & $E(B-V)$ & [Fe/H] & FU/FO \\ \hline

T Vel      & 0.667 &  9.802 $\pm$  0.285 & -2.050 & -2.690 & -3.370 & -4.260 & 0.281 & -0.02 $\pm$   0.08 & 0 \\  
$\delta$ Cep  & 0.730 &  7.165 $\pm$  0.193 & -2.950 & -3.510 & -4.140 & -4.890 & 0.092 &  0.03 $\pm$   0.06 & 0 \\  
Z Lac      & 1.037 & 11.637 $\pm$  0.262 & -3.860 & -4.560 & -5.290 & -6.090 & 0.404 &  0.01 $\pm$   0.08 & 0 \\  
BN Pup     & 1.136 & 12.950 $\pm$  0.239 & -3.760 & -4.510 & -5.270 & -6.180 & 0.438 &  0.01 $\pm$   0.06 & 0 \\  
RZ Vel     & 1.310 & 11.020 $\pm$  0.141 & -4.250 & -5.040 & -5.820 & -6.820 & 0.335 & -0.07 $\pm$   0.06 & 0 \\  
VZ Pup     & 1.365 & 13.083 $\pm$  0.271 & -4.320 & -5.010 & -5.720 & -6.560 & 0.471 & -0.16 $\pm$   0.08 & 0 \\  
RY Vel     & 1.449 & 12.019 $\pm$  0.155 & -4.690 & -5.500 & -6.300 & -7.280 & 0.562 & -0.03 $\pm$   0.09 & 0 \\  
AQ Pup     & 1.479 & 12.522 $\pm$  0.216 & -4.650 & -5.510 & -6.410 & -7.400 & 0.512 & -0.14 $\pm$   0.06 & 0 \\  
X Cyg      & 1.214 & 10.265 $\pm$  0.048 & -3.960 & -4.830 & -5.610 & -6.530 & 0.288 &  0.12 $\pm$   0.05 & 0 \\  
SW Vel     & 1.370 & 12.032 $\pm$  0.096 & -4.240 & -5.050 & -5.880 & -6.920 & 0.349 & -0.05 $\pm$   0.09 & 0 \\  
V340 Nor   & 1.053 & 11.152 $\pm$  0.124 & -2.990 & -3.830 & -4.690 & -5.680 & 0.315 & -0.09 $\pm$   0.08 & 0 \\  
S Nor      & 0.989 &  9.877 $\pm$  0.112 & -3.310 & -4.070 & -4.830 & -5.790 & 0.189 &  0.01 $\pm$   0.07 & 0 \\  
V Cen      & 0.740 &  9.175 $\pm$  0.138 & -2.710 & -3.300 & -3.960 & -4.770 & 0.289 & -0.05 $\pm$   0.06 & 0 \\  
CV Mon     & 0.731 & 11.021 $\pm$  0.151 & -2.490 & -3.070 & -3.830 & -4.680 & 0.714 & -0.04 $\pm$   0.06 & 0 \\  
$\eta$ Aql  & 0.856 &  7.132 $\pm$  0.095 & -3.080 & -3.730 & -4.420 & -5.310 & 0.149 &  0.06 $\pm$   0.06 & 0 \\  
T Mon      & 1.432 & 10.652 $\pm$  0.063 & -4.240 & -5.210 & -6.090 & -7.220 & 0.209 &  0.11 $\pm$   0.06 & 0 \\  
RS Pup     & 1.617 & 11.300 $\pm$  0.228 & -4.790 & -5.760 & -6.700 & -7.820 & 0.446 &  0.17 $\pm$   0.10 & 0 \\  
U Sgr      & 0.829 &  8.961 $\pm$  0.086 & -2.910 & -3.600 & -4.340 & -5.150 & 0.403 &  0.03 $\pm$   0.06 & 0 \\  
WZ Sgr     & 1.339 & 11.461 $\pm$  0.129 & -4.040 & -4.970 & -5.890 & -7.050 & 0.467 &  0.00 $\pm$   0.15 & 0 \\  
EV Sct     & 0.490 & 10.920 $\pm$  0.150 & -2.150 & -2.683 & -3.345 & -4.077 & 0.621 & -0.37 $\pm$   0.06 & 1 \\  
V1726 Cyg  & 0.651 & 12.690 $\pm$  0.150 & -2.716 & -3.338 & -3.986 &        & 0.548 & -0.02 $\pm$   0.14 & 0 \\  
CF Cas     & 0.688 & 12.690 $\pm$  0.150 & -2.541 & -3.209 & -3.901 & -4.859 & 0.531 & -0.10 $\pm$   0.07 & 0 \\  
QZ Nor     & 0.578 & 11.170 $\pm$  0.150 & -2.574 & -3.181 & -3.813 & -4.636 & 0.286 &  0.06 $\pm$   0.06 & 1 \\  
V367 Sct   & 0.799 & 11.320 $\pm$  0.150 & -2.921 & -3.543 & -4.323 & -5.020 & 1.208 & -0.01 $\pm$   0.11 & 0 \\  
DL Cas     & 0.903 & 11.220 $\pm$  0.150 & -3.071 & -3.743 & -4.435 & -5.444 & 0.479 &  0.02 $\pm$   0.10 & 0 \\  
TW Nor     & 1.033 & 11.470 $\pm$  0.150 & -2.906 & -3.697 & -4.498 & -5.443 & 1.214 &  0.03 $\pm$   0.06 & 0 \\  
KQ Sco     & 1.458 & 12.360 $\pm$  0.150 & -4.241 & -5.339 & -6.401 & -7.666 & 0.839 &  0.16 $\pm$   0.05 & 0 \\  
S Vul      & 1.838 & 13.240 $\pm$  0.150 & -5.588 & -6.740 & -7.803 & -8.862 & 0.737 & -0.02 $\pm$   0.05 & 0 \\  
W Sgr      & 0.881 &  7.876 $\pm$  0.123 & -2.945 & -3.571 & -4.230 &        & 0.110 & -0.01 $\pm$   0.07 & 0 \\  
$\beta$ Dor & 0.993 &  7.519 $\pm$  0.132 & -3.157 & -3.920 & -4.676 & -5.572 & 0.040 & -0.01 $\pm$   0.12 & 0 \\  
$\zeta$ Gem & 1.007 &  7.782 $\pm$  0.117 & -3.109 & -3.897 & -4.717 &        & 0.010 &  0.02 $\pm$   0.05 & 0 \\  
Y Oph      & 1.234 &  9.024 $\pm$  0.164 & -4.331 & -5.000 & -5.734 & -6.537 & 0.650 &  0.05 $\pm$   0.07 & 0 \\  
RX Aur     & 1.065 & 11.101 $\pm$  0.195 & -3.622 & -4.337 & -5.009 &        & 0.270 & -0.10 $\pm$   0.07 & 0 \\  
SZ Tau     & 0.498 &  8.725 $\pm$  0.133 & -2.623 & -3.151 & -3.761 & -4.501 & 0.290 &  0.04 $\pm$   0.06 & 1 \\  
EU Tau     & 0.323 & 10.269 $\pm$  0.159 & -2.258 & -2.737 & -3.310 &        & 0.170 & -0.06 $\pm$   0.06 & 1 \\  
T Vul      & 0.647 &  8.920 $\pm$  0.142 & -2.794 & -3.364 & -3.985 & -4.740 & 0.060 & -0.05 $\pm$   0.06 & 0 \\  
SV Vul     & 1.653 & 11.405 $\pm$  0.078 & -5.245 & -6.066 & -6.821 & -7.655 & 0.570 &  0.05 $\pm$   0.06 & 0 \\  

\hline
\end{tabular}
\label{TAB-GAL-P}
\end{table*}

\subsection{Magellanic Cloud cepheids}

The metallicity determination for 10 LMC and 6 SMC cepheids come from
Luck et al. (1998) (The values as derived using Kurucz ATLAS9 model
atmospheres are being used). An error bar of 0.10 dex had been adopted.

Optical photometry was taken from Martin \& Warren (1979), Freedman et
al. (1985), Moffett et al. (1998), Barnes et al. (1999), Caldwell et
al. (2001), and Sebo et al. (2002), and infrared photometry from Welch
et al. (1987) and Laney \& Stobie (1986b). Corresponding $BVIK$
datasets were joined (with offsets sometimes applied), and analysed
using the program ``Period98'' (Sperl 1998) to find (improved) periods
and mean magnitudes (and Fourier coefficients). The results are listed
in Table~\ref{TAB-MC-P}. The error quoted for the mean magnitudes is
the rms in the fit, but has been used as an error estimate.

Reddening values when available are taken from Caldwell \& Laney
(1994), Laney \& Stobie (1986a, 1994) and Caldwell \& Coulson (1985)
while for HV 900 and HV 909 an average of 0.06 has been adopted. The
effect of increased reddening will be discussed later.

The cepheids are not located all at the same distance because of depth
and projection effects. The magnitudes to be added to the observed
values to correct for this are listed in Table~\ref{TAB-MC-P} and have
been determined using the position angle and inclination from Van der
Marel \& Cioni (2001) for the LMC and Caldwell \& Laney (1991) for the
SMC. The effect is small but should be considered when aiming for the
highest accuracies.

\begin{table*}
\caption{Basic data of LMC/SMC cepheid sample with individual metallicity determinations }
\begin{tabular}{rrrrrcccc} \hline

Name      &    $P$   &        [Fe/H]      & $E(B-V)$ & depth   &  $m_{\rm V}$ & $m_{\rm B}$ & $m_{\rm I}$ &   $m_{\rm K}$  \\ \hline

HV 879    &   36.824 &  -0.56 $\pm$ 0.10  &  0.060  &  -0.030  &  13.401 $\pm$  0.036  &  13.516 $\pm$  0.060  &  12.330 $\pm$  0.022  &  11.103 $\pm$  0.020\\ 
HV 883    &  133.432 &  -0.45 $\pm$ 0.10  &  0.104  &  -0.005  &  12.161 $\pm$  0.057  &  13.363 $\pm$  0.121  &  11.014 $\pm$  0.029  &   9.720 $\pm$  0.027\\ 
HV 900    &   47.479 &  -0.38 $\pm$ 0.10  &  0.06   &  -0.030  &  12.832 $\pm$  0.012  &  13.853 $\pm$  0.018  &  11.848 $\pm$  0.018  &  10.706 $\pm$  0.054\\ 
HV 909    &   37.591 &  -0.28 $\pm$ 0.10  &  0.06   &  -0.036  &  12.805 $\pm$  0.023  &  13.652 $\pm$  0.031  &  11.939 $\pm$  0.021  &  10.880 $\pm$  0.011\\ 
HV 2257   &   39.390 &  -0.34 $\pm$ 0.10  &  0.061  &  -0.040  &  13.105 $\pm$  0.030  &  14.158 $\pm$  0.025  &  12.084 $\pm$  0.022  &  10.950 $\pm$  0.036\\ 
HV 2338   &   42.193 &  -0.44 $\pm$ 0.10  &  0.040  &  -0.057  &  12.823 $\pm$  0.043  &  13.824 $\pm$  0.061  &  11.851 $\pm$  0.023  &  10.730 $\pm$  0.030\\ 
HV 2447   &  118.338 &  -0.40 $\pm$ 0.10  &  0.043  &  +0.017  &  12.021 $\pm$  0.037  &  13.328 $\pm$  0.058  &  10.880 $\pm$  0.031  &   9.614 $\pm$  0.069\\ 
HV 2827   &   78.820 &  -0.24 $\pm$ 0.10  &  0.080  &  +0.094  &  12.324 $\pm$  0.023  &  13.538 $\pm$  0.035  &  11.196 $\pm$  0.019  &   9.850 $\pm$  0.033\\ 
HV 2883   &  108.923 &  -0.16 $\pm$ 0.10  &  0.015  &  +0.067  &  12.511 $\pm$  0.037  &  13.780 $\pm$  0.061  &  11.384 $\pm$  0.030  &  10.111 $\pm$  0.028\\
HV 5497   &   99.587 &  -0.48 $\pm$ 0.10  &  0.095  &  +0.029  &  11.940 $\pm$  0.036  &  13.169 $\pm$  0.043  &  10.817 $\pm$  0.039  &   9.475 $\pm$  0.023\\ 
\hline
HV 821    &  127.253 &  -0.84 $\pm$ 0.10  &  0.074  &  -0.073  &  11.997 $\pm$  0.037  &  13.025 $\pm$  0.054  &  10.936 $\pm$  0.036  &   9.747 $\pm$  0.043\\ 
HV 824    &   65.861 &  -0.94 $\pm$ 0.10  &  0.030  &  -0.037  &  12.401 $\pm$  0.026  &  13.288 $\pm$  0.028  &  11.486 $\pm$  0.024  &  10.370 $\pm$  0.020\\ 
HV 829    &   84.800 &  -0.61 $\pm$ 0.10  &  0.030  &  -0.018  &  11.957 $\pm$  0.021  &  12.796 $\pm$  0.018  &  11.053 $\pm$  0.021  &   9.952 $\pm$  0.036\\ 
HV 834    &   73.566 &  -0.59 $\pm$ 0.10  &  0.020  &  +0.006  &  12.254 $\pm$  0.044  &  13.102 $\pm$  0.068  &  11.365 $\pm$  0.039  &  10.230 $\pm$  0.034\\ 
HV 837    &   42.721 &  -0.91 $\pm$ 0.10  &  0.042  &  +0.022  &  13.288 $\pm$  0.021  &  14.202 $\pm$  0.023  &  12.322 $\pm$  0.018  &  11.178 $\pm$  0.023\\ 
HV 11157  &   69.491 &  -0.77 $\pm$ 0.10  &  0.080  &  +0.005  &  12.939 $\pm$  0.010  &  14.021 $\pm$  0.010  &  11.862 $\pm$  0.009  &  10.599 $\pm$  0.025\\ 

\hline
\end{tabular}
\label{TAB-MC-P}
\end{table*}

\section{The model and results}

The following Period-Luminosity-Colour-Metallicity ($PLCZ$)-relation
is fitted to the data, where both the zero point and slope are allowed to vary quadratically with metallicity:
\begin{displaymath}
M = \alpha_1  + \alpha_2 \log {\rm [Fe/H]}  + \alpha_3 (\log {\rm [Fe/H]})^2 
\end{displaymath}
\begin{displaymath}
\hspace{5mm} + \alpha_4 \; (1 + \frac{\alpha_5}{\alpha_4} \, \log {\rm [Fe/H]} + \frac{\alpha_6}{\alpha_4} \, (\log {\rm [Fe/H]})^2) \log P_0 
\end{displaymath}
\begin{displaymath}
\hspace{5mm} + \alpha_7 \; (1 + \frac{\alpha_8}{\alpha_7} \, \log {\rm [Fe/H]} + \frac{\alpha_9}{\alpha_7} \, (\log {\rm [Fe/H]})^2) (\log P_0)^2
\end{displaymath}
\begin{equation}
\hspace{5mm} + \beta (m_1 - m_2) + \Delta_{\rm LMC}\; x_{\rm LMC} + \Delta_{\rm SMC}\; x_{\rm SMC},
\end{equation}
where $M$ is the absolute magnitude in a given photometric band, $P_0$
the fundamental period (in days), [Fe/H] the metallicity on a
logarithmic scale relative to Solar, $(m_1 - m_2)$ represents an
arbitrary colour term, $x_{\rm LMC}$ and $x_{\rm SMC}$ are 1 for the
LMC, respectively SMC, cepheids and 0 otherwise. The absolute
magnitudes for the MC cepheids are calculated assuming default DM of
18.50 and 18.90, respectively (which represent ``rounded'' values that
are within 0.1 mag of recent determinations, see e.g. Feast 2003,
Walker 2003, Storm et al. 2004, and for the LMC corresponds to the
value adopted by Mould et al. and Freedman et al. in the {\sc hst} key
project), and $\Delta_{\rm LMC}$ and $\Delta_{\rm SMC}$ represent
corrections to that.

\begin{figure}
\centerline{\psfig{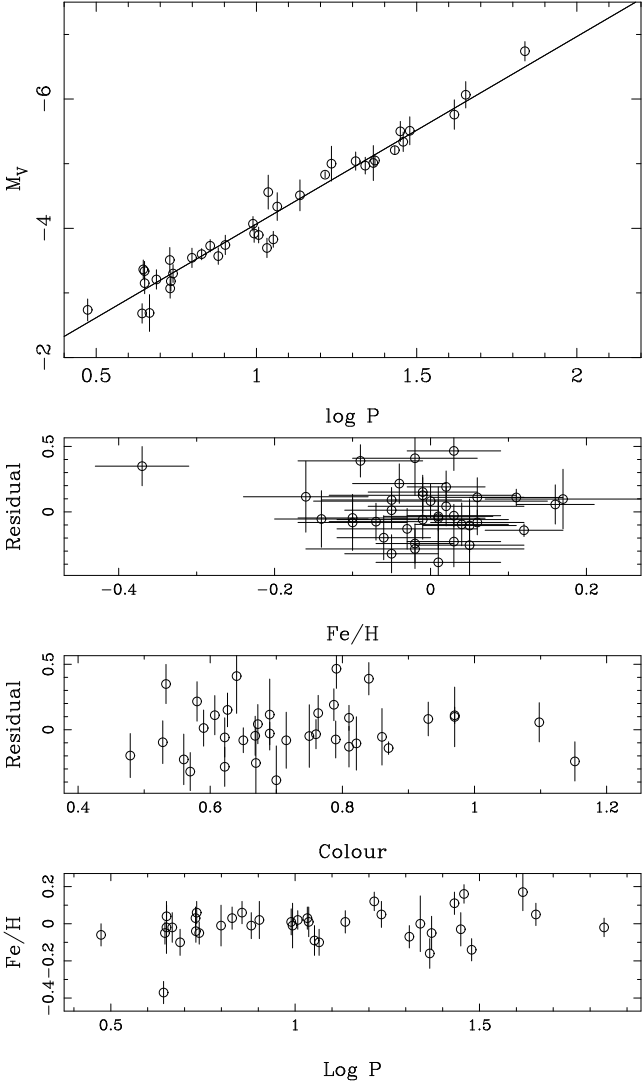}}
\caption[]{
Top: Period-Luminosity relation in the V-band for the Galaxy. Shown
are the data points and the best fitting linear $PL$-relation. The
middle panels shown the residuals plotted versus metallicity and
$(B-V)_0$ colour. The bottom panel shows [Fe/H] versus Period. Stars
are plotted at their fundamental period.
}
\label{PLV}
\end{figure}

\begin{figure}
\centerline{\psfig{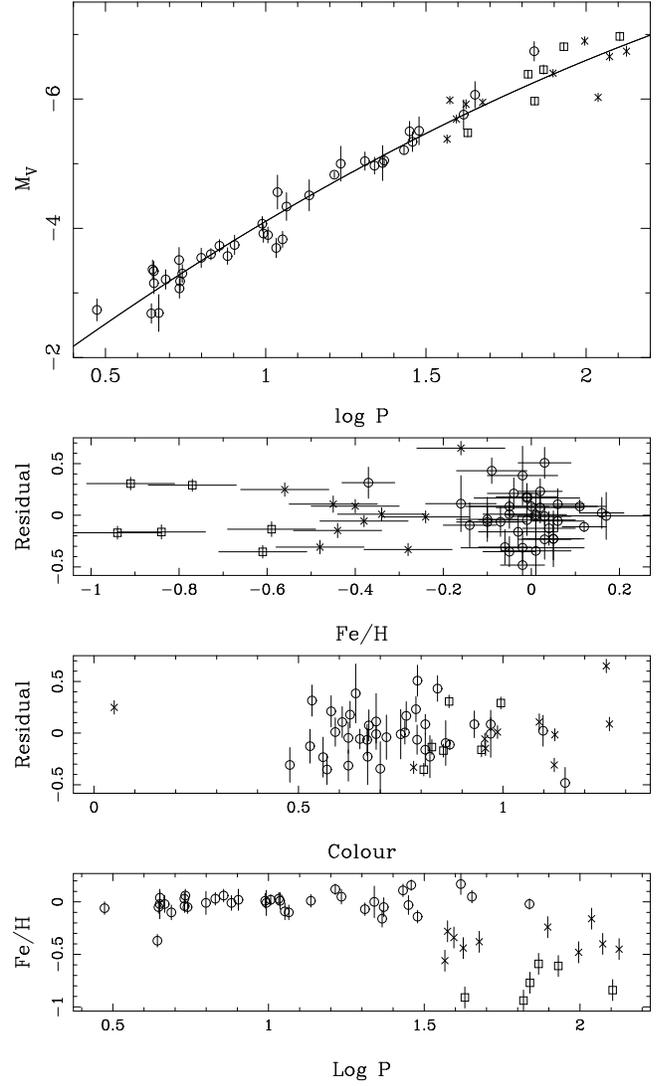}}
\caption[]{
Top: Period-Luminosity relation in the V-band showing data points
($\circ$ = Galactic, $\Box$ = LMC, x = SMC) and the fit allowing for a
quadratic term in $\log P$. Other panels as Fig.~1.
}
\label{PL-V1}
\end{figure}

\begin{figure}
\centerline{\psfig{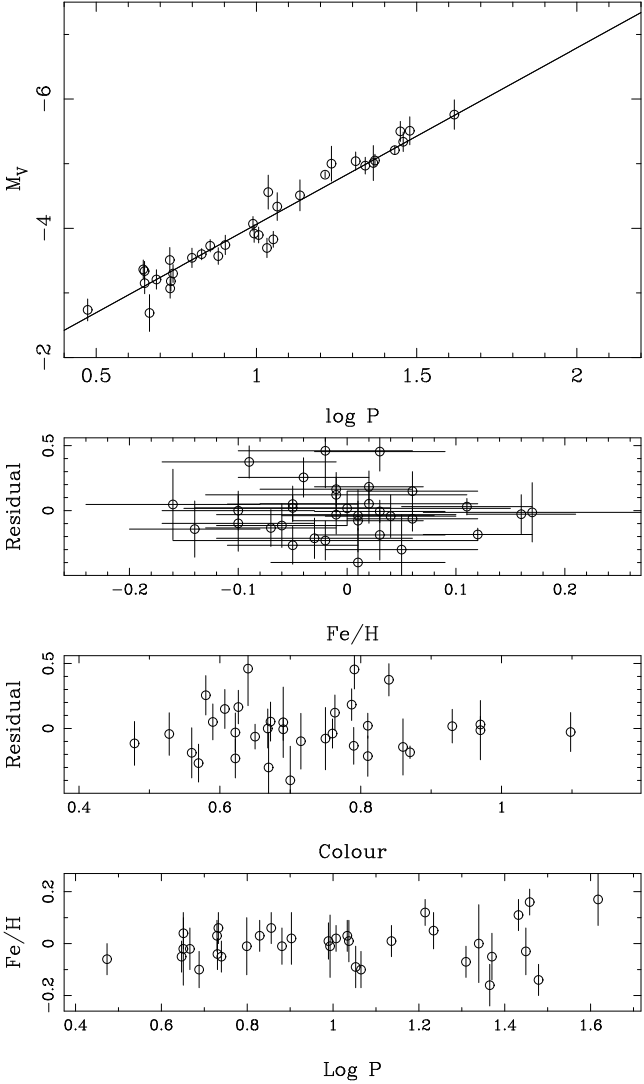}}
\caption[]{
Top: Linear Period-Luminosity relation in the V-band for the Galaxy,
with EV Sct excluded and for $\log P < 1.65$.
Other panels as Fig.~1.
}
\label{PL-V2}
\end{figure}

\begin{figure}
\centerline{\psfig{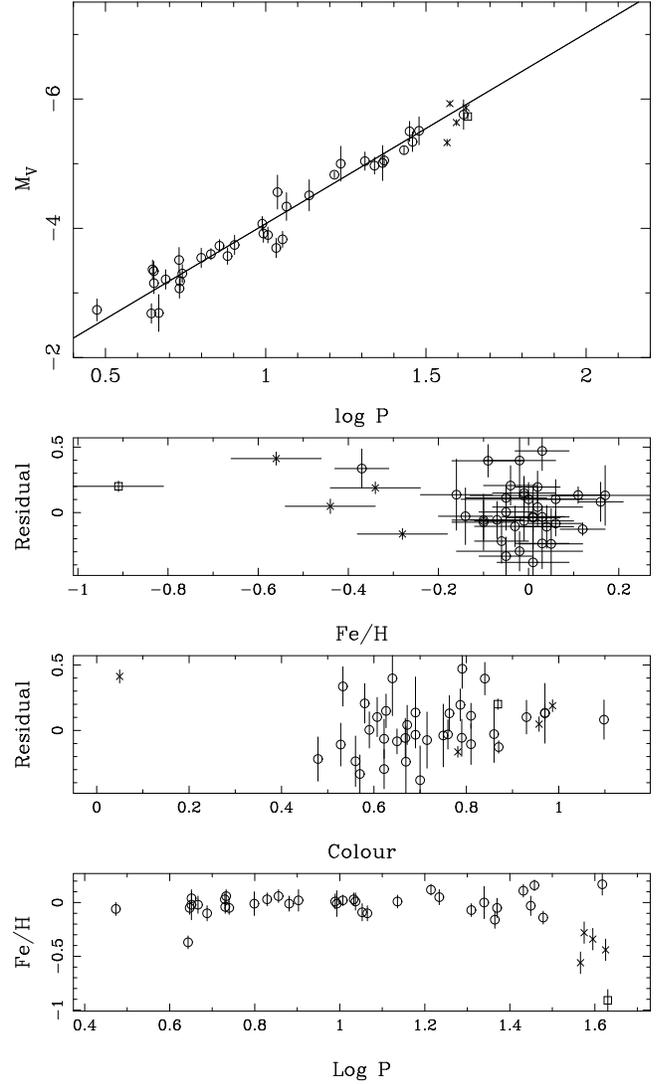}}
\caption[]{
Top: Linear Period-Luminosity relation in the V-band for Galaxy, SMC,
LMC for $\log P < 1.65$.  Other panels as Fig.~1.
}

\label{PL-V3}
\end{figure}

The observed periods of the first overtone pulsators are
``fundamentalised'' using $P_0 = P_1 / (0.716 - 0.027 \log P_1)$
(Feast \& Catchpole 1997).

The results of the fitting are listed in Table~\ref{TAB-RES1}. In the
fitting the error in the absolute magnitudes includes the error in the
distance modulus for the Galactic cepheids, a 0.01 error in $E(B-V)$
for all cepheids, the error in the mean magnitudes for the MC cepheids
and a 0.01 error in the depth effect for the MC cepheids. The
(non-negligible) error in the fit when including a metallicity effect,
due to the error therein, will be discussed in detail below.

First, the classical (linear) PL-relations in $BVIWK$ are determined
for stars with $\log P > 0.4$ for the Galaxy only, and the fit for the
$V$-band is shown in Fig.~1 as an example. The slopes in $BVIWK$,
respectively, $BVI$ agree at the 2$\sigma$ or better level with those
in F03 and T03 (his mean relation). At $\log P = 1$ the agreement with
F03 for the absolute magnitudes is excellent in all filters (typically
0.03 mag or about 1$\sigma$ or better), with T03 the difference is
larger (about 0.1 mag fainter, which is also at the 1$\sigma$
level). The reason for the latter effect is probably that in 10 out of
15 cepheids in Table~\ref{TAB-GAL-D} the adopted distance is in the end smaller
than the one taken by T03. 

Second, the PL-relations in $BVIWK$ are determined for all three
galaxies combined and allowing for the quadratic term in $\log P$ but
not yet for any metallicity dependence. The result in the $V$-band is
shown in Fig.~2. This fit and the results in the tables show that
there is a significant quadratic term in $\log P$ when the long-period
MC cepheids are included in the fit. 

One can also observe that the best fitting distance to, in particular,
the SMC seems to have become shorter than the default value of 18.90
(there is some scatter depending on the band, and some dependence on
the reddening) and that the error bar in the linear term in $\log P$
has increased significantly.

Deviations from a linear $PL$-relation have long been observed,
e.g. Sandage \& Tammann (1968) and Martin et al. (1979) at $\log P
\approx 1.7-1.8$.  The presence of a quadratic term in the
$PL$-relation was predicted by the model calculations of Bono et
al. (1999) in the V-band (see also Fiorentino et al. 2002), and they
proposed two linear relations with the break at $\log P$ = 1.4. Our
results show that the linear relation may be valid to slightly longer
periods than $\log P = 1.4$.

The upper cut-off in $\log P$ is determined empirically in such a way
that the quadratic term becomes statistically insignificant (in all
colours $BVIWK$), and a linear fit in $\log P$ is therefore
justified. This cut-off is $\log P = 1.65$, in good agreement with
the quoted earlier observational results.

In view of this, two samples will be considered, both with cepheids in
the period range $0.4 < \log P_0 < 1.65$, viz. a purely Galactic
sample, and a second sample that includes all three galaxies but only
1 SMC and 4 LMC stars. For this reason $\Delta_{\rm LMC}$ and
$\Delta_{\rm SMC}$ are no longer fitted in what follows.

Next the effect of metallicity is considered for the Galactic sample,
as a possible change of the zero point. As is clear from
Fig.~\ref{PLV}, any dependence would be heavily influenced by the
outlier EV Sct which has [Fe/H] = $-0.37$.  Therefore
Table~\ref{TAB-RES1} includes both cases, where this star is either
kept or is excluded from the sample when fitting the $PLZ$-relation in
the Galaxy. The coefficient $\alpha_2$ is strongly negative ($-0.2$ to
$-0.8$ mag/dex) but only significant at the 2-3 $\sigma$ level
considering the formal error in the fit. Figure~\ref{PL-V2} shows the
result for the $V$-band, excluding EV Sct.

As noted earlier, there are some worrisome discrepancies between
independent metallicity determinations for the same object (see
Table~2). The fit parameters quoted are based on assuming that there
is no error in the independent variables period and metallicity. For
the period this is certainly the case but not for the metallicity.
Therefore a set of 1000 simulations was run where the metallicity of
each star was set to its observed value plus its attributed error
multiplied by a number drawn from a Gaussian distribution. The fitting
was then repeated and the 23th and 977th ordered values, which
correspond to $\pm 2 \sigma$, of the fit parameters determined. The
analysis of the $VWK$ bands shows that the $\sigma$ due to the error
in the metallicity is $\approx$0.03 in both slope and zero point. This
error should be quadratically added to the errors in the fit listed in
Table~\ref{TAB-RES1}, but this has only a marginal effect. On the
other hand, the $\sigma$ due to the error in the metallicity
$\alpha_2$ for $VWK$ with (without) EV Sct included in the sample is
0.18 (0.24), 0.16 (0.21) and 0.14 (0.17), respectively. The total
error in $\alpha_2$ becomes, respectively, 0.31 (0.39), 0.30 (0.38),
0.29 (0.36) mag/dex. This implies that the metallicity effect in the
zero point for the Galactic sample is $-0.6 \pm 0.4$ mag/dex in $VWK$
when EV Sct is excluded and $-0.8 \pm 0.3$ when EV Sct is included.

These results indicate that to obtain a better estimate of the
metallicity effect based on Galactic cepheids alone, one would need
more data points at low and high metallicities. Good candidates would
be cepheids in the direction of the centre and the anti-centre of the
Galaxy. However, these stars would be at considerable distance and
probably an enlargement of the Galactic sample would be limited by our
ability to accurately determine individual distances.

When all three galaxies are considered, the baseline in metallicity is
now much larger (EV Sct is included, but the star does not influence
the solution just because the baseline in metallicty is much larger
now). The coefficient $\alpha_2$ is now determined at the $\sim$6
$\sigma$ level. In the $B$-band its value is slightly positive, in
$VIWK$ negative at about $-0.27$ mag/dex. Figure~\ref{PL-V3} shows the
result for the $V$-band. The error bars listed have to be increased to
include the uncertainty due to the error in the metallicity itself. As
before, Monte Carlo simulations have been performed and $\sigma$
values of 0.061, 0.035, 0.050, respectively, in $VWK$ were determined
in $\alpha_2$. This implies that the metallicity effect is $-0.27 \pm
0.08$ mag/dex, and is within the error the same in $VWK$. Error bars
of 0.04 should be added quadratically to the listed ones for the
slopes and zero points (independent of photometric band) due to the
error in the metallicity.

The results in $B$ and to a lesser extent in $V$ and $I$ depend
somewhat on the adopted reddenings to the MC cepheids. The last block
of results in Table~\ref{TAB-RES1} refers to the situation where a
uniform reddening of $E(B-V)$ = 0.151 and 0.092 for LMC and SMC,
respectively, has been adopted (the mean reddening of cepheids adopted
by the OGLE team). The results in the Wesenheit index and the $K$-band
are essentially unchanged.

When considering the metallicity effect in Galaxy, LMC and SMC it was
sofar assumed that the slope in the $\log P$ dependence does not
depend on metallicity. In reality, there is increasing evidence that
the slopes in the Galaxy, LMC and SMC are different, see e.g. the
recent discussion in Fouqu\'e et al. (2003). Equation~(1) allows for
an explicit metallicity dependence of the slope in $\log P$, and a
test calculation allowing $\alpha_5$ to vary was performed. A
non-conclusive (1$\sigma$) result was obtained and this is likely due
to the very small numbers of MC cepheids. When an increased sample of
MC cepheids with metallicity determinations becomes available the
metallicity dependence of the slope in $\log P$ can be investigated in
more detail.

More complicated fitting formulae, for example with a quadratic
metallicity dependence, or a colour term, have been tried but did not
result in parameters which are determined in a significant way.

\begin{table*}
\caption{Results of the fitting. Coefficients refer to Eq.~(1). The
last column gives the rms in the fit}
\begin{tabular}{cccccccc} \hline
  &  $\alpha_1$         & $\alpha_4$           & $\alpha_7$           &  $\alpha_2$        &  $\Delta_{\rm LMC}$ & $\Delta_{\rm SMC}$ & $\sigma$  \\
\hline
\multicolumn{8}{c}{$PL$-relation Galaxy} \\
B & -0.881 $\pm$  0.080 &  -2.444 $\pm$  0.070 &  0.000 $\pm$  0.000 & 0.000 $\pm$  0.000 &   0.000 $\pm$  0.000 &  0.000 $\pm$  0.000 & 0.234 \\
V & -1.167 $\pm$  0.078 &  -2.901 $\pm$  0.069 &  0.000 $\pm$  0.000 & 0.000 $\pm$  0.000 &   0.000 $\pm$  0.000 &  0.000 $\pm$  0.000 & 0.200 \\
I & -1.644 $\pm$  0.076 &  -3.174 $\pm$  0.067 &  0.000 $\pm$  0.000 & 0.000 $\pm$  0.000 &   0.000 $\pm$  0.000 &  0.000 $\pm$  0.000 & 0.188 \\
W & -2.346 $\pm$  0.082 &  -3.637 $\pm$  0.072 &  0.000 $\pm$  0.000 & 0.000 $\pm$  0.000 &   0.000 $\pm$  0.000 &  0.000 $\pm$  0.000 & 0.216 \\
K & -2.292 $\pm$  0.080 &  -3.424 $\pm$  0.067 &  0.000 $\pm$  0.000 & 0.000 $\pm$  0.000 &   0.000 $\pm$  0.000 &  0.000 $\pm$  0.000 & 0.197 \\
\multicolumn{8}{c}{$PL$-relation Galaxy + LMC + SMC; all periods} \\
B & -0.425 $\pm$  0.142 &  -3.427 $\pm$  0.210 &  0.481 $\pm$  0.073 & 0.000 $\pm$  0.000 &  -0.137 $\pm$  0.047 & -0.530 $\pm$  0.052 & 0.289 \\
V & -0.706 $\pm$  0.134 &  -3.857 $\pm$  0.190 &  0.455 $\pm$  0.061 & 0.000 $\pm$  0.000 &   0.054 $\pm$  0.042 & -0.251 $\pm$  0.045 & 0.235 \\
I & -1.091 $\pm$  0.127 &  -4.295 $\pm$  0.176 &  0.522 $\pm$  0.053 & 0.000 $\pm$  0.000 &   0.053 $\pm$  0.038 & -0.173 $\pm$  0.040 & 0.216 \\
W & -2.020 $\pm$  0.164 &  -4.341 $\pm$  0.263 &  0.345 $\pm$  0.103 & 0.000 $\pm$  0.000 &   0.132 $\pm$  0.065 &  0.015 $\pm$  0.074 & 0.216 \\
K & -1.365 $\pm$  0.138 &  -5.096 $\pm$  0.184 &  0.700 $\pm$  0.054 & 0.000 $\pm$  0.000 &   0.051 $\pm$  0.034 & -0.127 $\pm$  0.037 & 0.205 \\
\multicolumn{8}{c}{$PLZ$-relation Galaxy; $\log P < 1.65$; EV Sct excluded} \\
B & -1.005 $\pm$  0.091 &  -2.323 $\pm$  0.086 &  0.000 $\pm$  0.000 & -0.212 $\pm$  0.310 &  0.000 $\pm$  0.000 &  0.000 $\pm$  0.000 & 0.234 \\
V & -1.330 $\pm$  0.090 &  -2.731 $\pm$  0.085 &  0.000 $\pm$  0.000 & -0.634 $\pm$  0.309 &  0.000 $\pm$  0.000 &  0.000 $\pm$  0.000 & 0.198 \\
I & -1.799 $\pm$  0.089 &  -3.013 $\pm$  0.084 &  0.000 $\pm$  0.000 & -0.638 $\pm$  0.308 &  0.000 $\pm$  0.000 &  0.000 $\pm$  0.000 & 0.178 \\
W & -2.536 $\pm$  0.094 &  -3.441 $\pm$  0.088 &  0.000 $\pm$  0.000 & -0.653 $\pm$  0.313 &  0.000 $\pm$  0.000 &  0.000 $\pm$  0.000 & 0.188 \\
K & -2.301 $\pm$  0.097 &  -3.420 $\pm$  0.089 &  0.000 $\pm$  0.000 & -0.574 $\pm$  0.315 &  0.000 $\pm$  0.000 &  0.000 $\pm$  0.000 & 0.168 \\
\multicolumn{8}{c}{$PLZ$-relation Galaxy; $\log P < 1.65$; EV Sct included} \\
B & -1.002 $\pm$  0.091 &  -2.314 $\pm$  0.086 &  0.000 $\pm$  0.000 & -0.467 $\pm$  0.254 &  0.000 $\pm$  0.000 &  0.000 $\pm$  0.000 & 0.236 \\
V & -1.328 $\pm$  0.090 &  -2.725 $\pm$  0.085 &  0.000 $\pm$  0.000 & -0.792 $\pm$  0.253 &  0.000 $\pm$  0.000 &  0.000 $\pm$  0.000 & 0.199 \\
I & -1.798 $\pm$  0.089 &  -3.007 $\pm$  0.084 &  0.000 $\pm$  0.000 & -0.784 $\pm$  0.252 &  0.000 $\pm$  0.000 &  0.000 $\pm$  0.000 & 0.178 \\
W & -2.534 $\pm$  0.094 &  -3.437 $\pm$  0.087 &  0.000 $\pm$  0.000 & -0.782 $\pm$  0.255 &  0.000 $\pm$  0.000 &  0.000 $\pm$  0.000 & 0.186 \\
K & -2.293 $\pm$  0.097 &  -3.417 $\pm$  0.089 &  0.000 $\pm$  0.000 & -0.776 $\pm$  0.255 &  0.000 $\pm$  0.000 &  0.000 $\pm$  0.000 & 0.167 \\
\multicolumn{8}{c}{$PLZ$-relation Galaxy + LMC + SMC; $\log P < 1.65$; no fit to MC distances } \\
B & -0.807 $\pm$  0.073 &  -2.534 $\pm$  0.061 &  0.000 $\pm$  0.000 & +0.053 $\pm$  0.064 &  0.000 $\pm$  0.000 &  0.000 $\pm$  0.000 & 0.239 \\
V & -1.123 $\pm$  0.071 &  -2.948 $\pm$  0.058 &  0.000 $\pm$  0.000 & -0.274 $\pm$  0.057 &  0.000 $\pm$  0.000 &  0.000 $\pm$  0.000 & 0.196 \\
I & -1.603 $\pm$  0.066 &  -3.215 $\pm$  0.052 &  0.000 $\pm$  0.000 & -0.274 $\pm$  0.047 &  0.000 $\pm$  0.000 &  0.000 $\pm$  0.000 & 0.174 \\
W & -2.422 $\pm$  0.079 &  -3.558 $\pm$  0.067 &  0.000 $\pm$  0.000 & -0.237 $\pm$  0.088 &  0.000 $\pm$  0.000 &  0.000 $\pm$  0.000 & 0.182 \\
K & -2.193 $\pm$  0.070 &  -3.519 $\pm$  0.051 &  0.000 $\pm$  0.000 & -0.299 $\pm$  0.041 &  0.000 $\pm$  0.000 &  0.000 $\pm$  0.000 & 0.164 \\
\multicolumn{8}{c}{$PLZ$-relation Galaxy + LMC + SMC; $\log P < 1.65$; no fit to MC distances; increased reddening } \\
B & -0.552 $\pm$  0.073 &  -2.812 $\pm$  0.061 &  0.000 $\pm$  0.000 & +0.266 $\pm$  0.064 &  0.000 $\pm$  0.000 &  0.000 $\pm$  0.000 & 0.273 \\
V & -0.885 $\pm$  0.070 &  -3.209 $\pm$  0.058 &  0.000 $\pm$  0.000 & -0.164 $\pm$  0.057 &  0.000 $\pm$  0.000 &  0.000 $\pm$  0.000 & 0.226 \\
I & -1.406 $\pm$  0.066 &  -3.432 $\pm$  0.052 &  0.000 $\pm$  0.000 & -0.267 $\pm$  0.047 &  0.000 $\pm$  0.000 &  0.000 $\pm$  0.000 & 0.196 \\
W & -2.424 $\pm$  0.079 &  -3.556 $\pm$  0.067 &  0.000 $\pm$  0.000 & -0.242 $\pm$  0.088 &  0.000 $\pm$  0.000 &  0.000 $\pm$  0.000 & 0.181 \\
K & -2.155 $\pm$  0.070 &  -3.560 $\pm$  0.051 &  0.000 $\pm$  0.000 & -0.304 $\pm$  0.041 &  0.000 $\pm$  0.000 &  0.000 $\pm$  0.000 & 0.165 \\

\hline
\end{tabular}
\label{TAB-RES1}
\end{table*}

\section{Discussion}

An attempt has been made, using the currently available data of
cepheids with direct metallicity determinations, to quantify the
metallicity dependence of the cepheid $PL$-relation. 

For a purely Galactic sample, the range in metallicity covered is too
small to draw any firm conclusions. A metallicity effect of $-0.6
\pm 0.4$ or $-0.8 \pm 0.3$ mag/dex in $VWK$ is derived, depending on
whether the only Galactic star with a significantly sub-solar
metallicity is excluded or included.

For the combined sample of Galactic, SMC and LMC cepheids the problem
is that most of the MC cepheids have such long periods, in the regime
where the $PL$-relation is no longer linear. Restricting the sample in
periods to $\log P < 1.65$ a metallicity effect of about $-0.27 \pm
0.08$ mag/dex in the zero point is found (in $VIWK$). However, the
sample of MC cepheids is presently too small to additionaly solve for
a metallicity dependence of the slope in $\log P$. Formally, the
result for the Galactic sample is in agreement with that comprising
all three galaxies.

This result based on the sample including the MC cepheids is in
agreement with other recent empirical estimates (most recently by
Kennicutt et al. 2003 and Storm et al. 2004). However, it is stressed
that this study is the first to use {\em direct} cepheid metallicity
determinations to arrive at this result. In agreement with Storm
et al. (2004) we find that in the range $V$ to $K$ the metallicity
effect seems not to depend on wavelength, contrary to theoretical
calculations (e.g. Bono et al. 1999), which predicted a decreasing
effect towards longer wavelengths. 

The method here developed (Eq.~(1)) is general and can easily be
extended to other functional forms, or to include other galaxies. In
principle, the distance to a galaxy and the metallicity dependence can
be derived independently. For the present restrictive sample this is
not the case however, since the distribution over metallicity is
correlated with the parent galaxy. In other words, $\alpha_2$,
$\Delta_{\rm LMC}$ and $\Delta_{\rm SMC}$ are strongly correlated: a
change in the DM to the MCs can be 'compensated for' by a change in
the metallicity effect. Therefore, it was decided to not solve for the
distance moduli. However, in the second block of results in
Table~\ref{TAB-RES1} (no metallicity effect, but allowing for a
quadratic term in $\log P$) there is some indication that the best
fitting DM might be slightly longer than 18.50 for the LMC and
slightly shorter than 18.90 for the SMC (although there is some
scatter depending on the wavelength). Simply adopting values of 18.55
and 18.80, respectively, within the error bars of the current best
independent estimates (Feast 2003, Walker 2003), would result in a
value for $\alpha_2$ of $-0.38 \pm 0.08$ mag/dex in $V$, that is, it
would make the metallicity effect slightly stronger (and similarly for
the other bands).

It is shown that the error in the independent variable metallicity is
a significant contributor to the final error in the metallicity effect
as it is larger than the error in the fitting in the dependent
variable, but of little effect in the final error in slope and
zero point. The mean error in metallicity of the stars in the Galactic
plus MC sample is about 0.08 dex. By repeating the Monte Carlo
simulations it was derived that for a uniform error bar of 0.05 dex the
two sources of error become comparable, and that for a uniform
error bar of 0.03 dex or less the error in the metallicity becomes
insignificant compared to the fit error.

Although the present results seem to indicate a significant
metallicity effect, more data are needed to quantify this effect
better. An important issue is an internally consistent metallicity
scale. The simulations even suggest an accuracy to a level of 0.05 dex
or better. However, this seems very difficult to achieve in practice
as the error in the metallicity not only depends on the S/N in the
spectra and the number of lines, but also on the uncertainties in the
adopted effective temperature, gravity and model atmospheres (e.g. Fry
\& Carney 1997). This implies that an improvement in the present
result must come from an increased sample of stars with periods
$\less\ 45 d$ spread over a range in metallicities as large as possible.

\acknowledgements{
This research has made use of the SIMBAD database, operated at CDS,
Strasbourg, France.
}

{}

\end{document}